\documentclass[a4paper]{jpconf}
\usepackage{graphicx}
\usepackage[font=footnotesize, caption=false]{subfig}

\begin{document}
\title{Algorithms and parameters for improved accuracy in physics data libraries}

\author{M Bati\v{c}$^{1,2}$, M Han$^3$, S Hauf$^4$, G Hoff$^{1,5}$, C H Kim$^3$, M Kuster$^6$,
M~G~Pia$^1$, P Saracco$^1$ and H Seo$^7$}

\address{$^1$ INFN Sezione di Genova, Genova 16146, Italy}
\address{$^2$ Jozef Stefan Institute, 1000 Ljubljana, Slovenia}
\address{$^3$ Department of Nuclear Engineering, Hanyang University, Seoul 133-791, Korea}
\address{$^4$ Technische Universit\"at Darmstadt, IKP, Germany}
\address{$^5$ Pontificia Universidade Catolica do Rio Grande do Sul, Porto Alegre, Brazil.}
\address{$^6$ European XFEL GmbH, Hamburg, Germany}
\address{$^7$Korea Atomic Energy Research Institute, Daejeon 305-353, Korea}

\ead{Maria.Grazia.Pia@cern.ch}

\begin{abstract}
Recent efforts for the improvement of the accuracy of physics data libraries
used in particle transport are summarized.
Results are reported about a large scale validation analysis of atomic
parameters used by major Monte Carlo systems (Geant4, EGS, MCNP, Penelope etc.);
their contribution to the accuracy of simulation observables is documented.
The results of this study motivated the development of a new atomic data
management software package, which optimizes the provision of state-of-the-art
atomic parameters to physics models.
The effect of atomic parameters on the simulation of radioactive decay is
illustrated. Ideas and methods to deal with physics models applicable to
different energy ranges in the production of data libraries, rather than at
runtime, are discussed.

\end{abstract}

\section{Introduction}
Physics data libraries play an important role in Monte Carlo simulation: they
provide fundamental atomic and nuclear parameters, and tabulations of basic
physics quantities used in particle transport, such as cross sections,
correction factors, secondary particle energies and angular distributions.
They contribute significantly to the accuracy of simulation results.

This paper reviews ongoing activities to improve the accuracy of physics data
libraries relevant to Geant4 \cite{g4nim,g4tns}, and novel ideas concerning the
role that data libraries could play for improved computational performance and
simplified management of physics models.

Due to the constraints imposed by the license governing the publication of
conference proceedings, it summarizes the main features and results of the
research in progress; extensive details and the full set of results are
meant to be included in dedicated publications in scholarly journals.

\section{Optimization of atomic parameters for particle transport}
\label{atomic}

The simulation of particle interactions with matter involves several atomic
parameters, whose values affect the physics models used for particle
transport and experimental observables calculated by the simulation.
Despite the fundamental character of these parameters, a consensus has not
always been achieved about their values, and different Monte Carlo codes use
different sets of parameters.

An extensive study is in progress to evaluate the effects on simulation accuracy 
due to a variety of compilations of atomic parameters.
The first cycle of this project \cite{binding_tns} has evaluated several collections of atomic
binding energies, including those used by major Monte Carlo systems:

\begin{itemize}
\item the compilation by Bearden and Burr \cite{bearden}, used by ISICS \cite{isics},
\item the compilation by Carlson \cite{carlson}, used by MCNP \cite{mcnp,mcnpx} and Penelope \cite{penelope},
\item the tabulation included in Evaluated Atomic Data Library (EADL) \cite{eadl}, used by Geant4 \cite{g4nim,g4tns},
\item the compilation assembled by Sevier in 1979 \cite{sevier1979}, used by GUPIX \cite{gupix},
\item the compilations included in the seventh and eighth editions of the Table
of Isotopes \cite{toi1978,toi1996}, respectively published in 1978 and 1996, and used by EGSnrc \cite{egsnrc} and EGS5 \cite{egs5},
\item the compilation by Williams included in the X-ray Data Booklet \cite{xbook}.
\end{itemize}

\begin{figure}
\centerline{\includegraphics[angle=0,width=12cm]{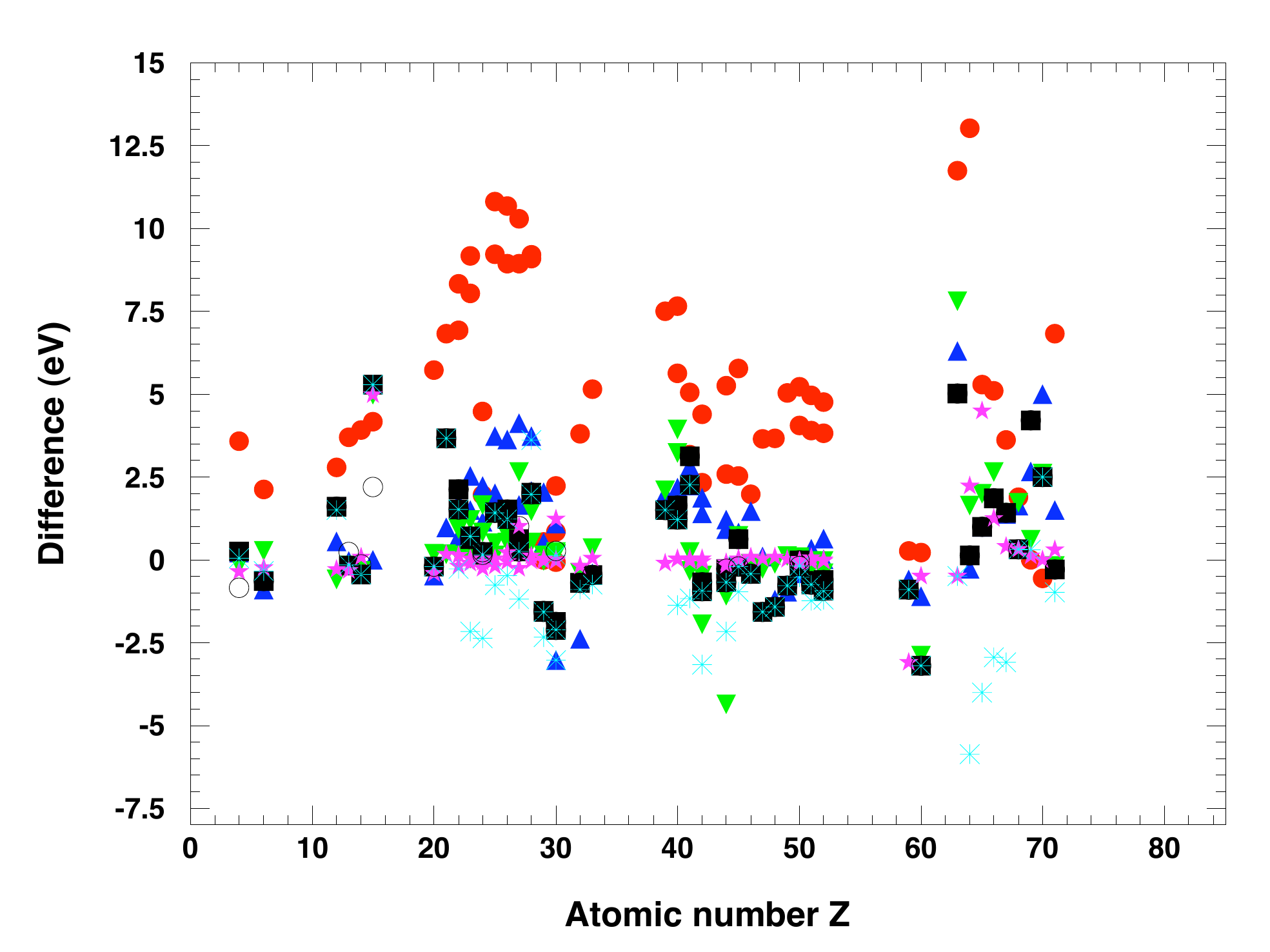}}
\caption{Difference between binding energies in various compilations and
NIST reference data versus atomic number: EADL (red circles),
Carlson (blue up triangles), Table of Isotopes 1996 (black squares), Table of
Isotopes 1978 (green down triangles), Williams (pink stars), Sevier
1979 (turquoise asterisks).}
\label{fig_principal}
\end{figure}

Figure \ref{fig_principal} shows the difference between the binding energies of
the various compilations and values recommended by NIST \cite{nist_xps}.
Discrepancies with respect to experimental values, and across the various 
compilations are visible.

\begin{figure}
\centerline{\includegraphics[angle=0,width=12cm]{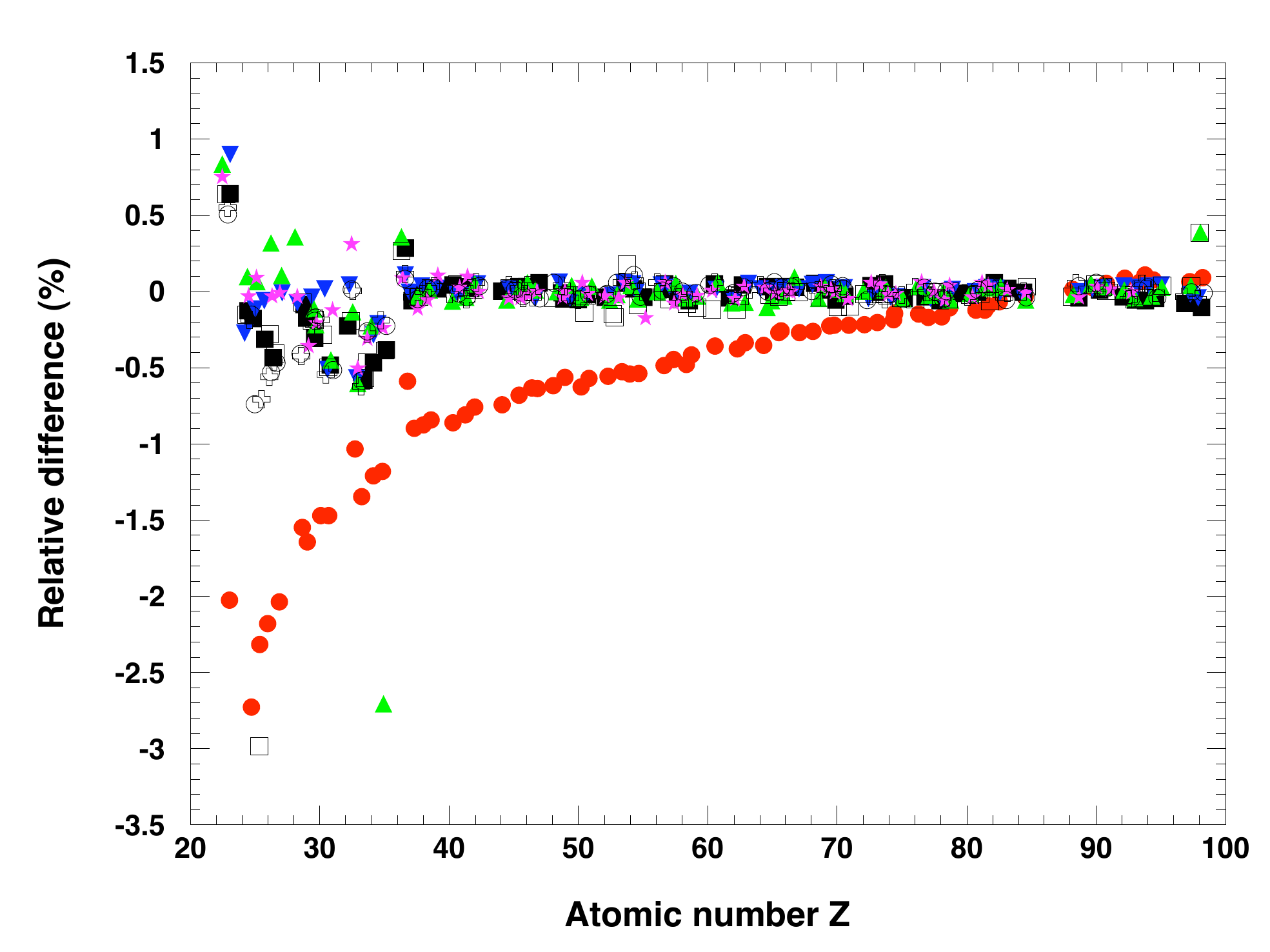}}
\caption{L$_1$M$_2$ transition, 
relative difference between X-ray energies resulting from atomic binding
energies in various compilations and experimental data from \cite{deslattes},
versus atomic number: EADL (red circles), Carlson (blue up triangles), Table of
Isotopes 1996 (black squares), Table of Isotopes 1978 (green down triangles),
Williams (pink stars), Sevier
1979 (turquoise asterisks) and G4AtomicShells (empty squares).}
\label{fig_l1m2}
\end{figure}

The atomic parameters used in the implementation of physics models affect
the accuracy of the simulation. 
An example is illustrated in figure \ref{fig_l1m2}, which shows the relative
difference between simulated and measured \cite{deslattes} fluorescence X-ray energies resulting
from $L_1M_2$ transitions produced by various compilations of atomic binding
energies.
EADL is responsible for the worst simulation accuracy in this use case, while the
other compilations exhibit similar behavior.

The validation test has demonstrated that no single compilation of atomic
binding energies is ideal for all applications.
A software package has been developed to manage atomic parameters needed by
Geant4 physics models; it allows simulations with multiple instances of atomic
physics objects to optimize the accuracy of physics processes sensitive to the
effect of atomic parameters.

\section{Improvements to Geant4 radioactive decay simulation}

The simulation of radioactive decays, as currently implemented in Geant4
\cite{rdm_chep2000,rdm_radecs2000}, exploits various data compilations:
\begin{itemize}
\item the Evaluated Nuclear Structure Data Files (ENSDF) \cite{ensdf} to
determine the decay type, decay emission and daughter nucleus resulting from the
transmutation of a given parent, and for handling photon evaporation;
\item compilations of electron conversion probabilities \cite{Band1976433,
Roesel197891, Hager19681};
\item the Evaluated Atomic Data Library (EADL) for atomic parameters involved in
the simulation of fluorescence and Auger electron emission.
\end{itemize}

Preliminary results of experimental validation tests of the current
implementation, assessing the accuracy of Geant4 simulations based on these
data, are reported in \cite{rdm_nss2009,rdm_mc2010}.
In parallel to the validation process, a major effort has been invested into
improving the software design, the physical accuracy and the computational
performance of Geant4 radioactive decay simulation.
The related developments and final validation results are extensively documented
in two forthcoming articles; only a few highlights concerning physics data
libraries used by the new code are summarized here.

The new code for radioactive decay simulation utilizes the latest version of ENSDF
for radioactive decay and photon evaporation data, and Bearden and Burr's 
compilation of atomic binding energies.
The software was subject to verification and validation.

The verification process involved a series of Geant4 simulations, each one
consisting of $10^{6}$ decays of an unstable nucleus in an otherwise empty
geometry.
The kinetic energy of the decay products was recorded in histograms separately
for each radiation type (electrons, photons and $\alpha$ particles).
In case of discrete emission contained in a single energy bin the intensity of
the emission is given by
\begin{equation}
\mathrm{Intensity} = \frac{\mathrm{Events\;in\;bin}}{\mathrm{Number\;of\;simulated\;decays}}
\end{equation}
An appropriate algorithm handled emissions distributed into multiple neighboring
bins.

Energy and intensity deviations of the simulated data with respect to ENSDF data,
that originate from experimental measurements, were calculated as part of the
software verification process.

\begin{figure}[!htbp]
\centerline{\subfloat{\includegraphics[width=8cm]{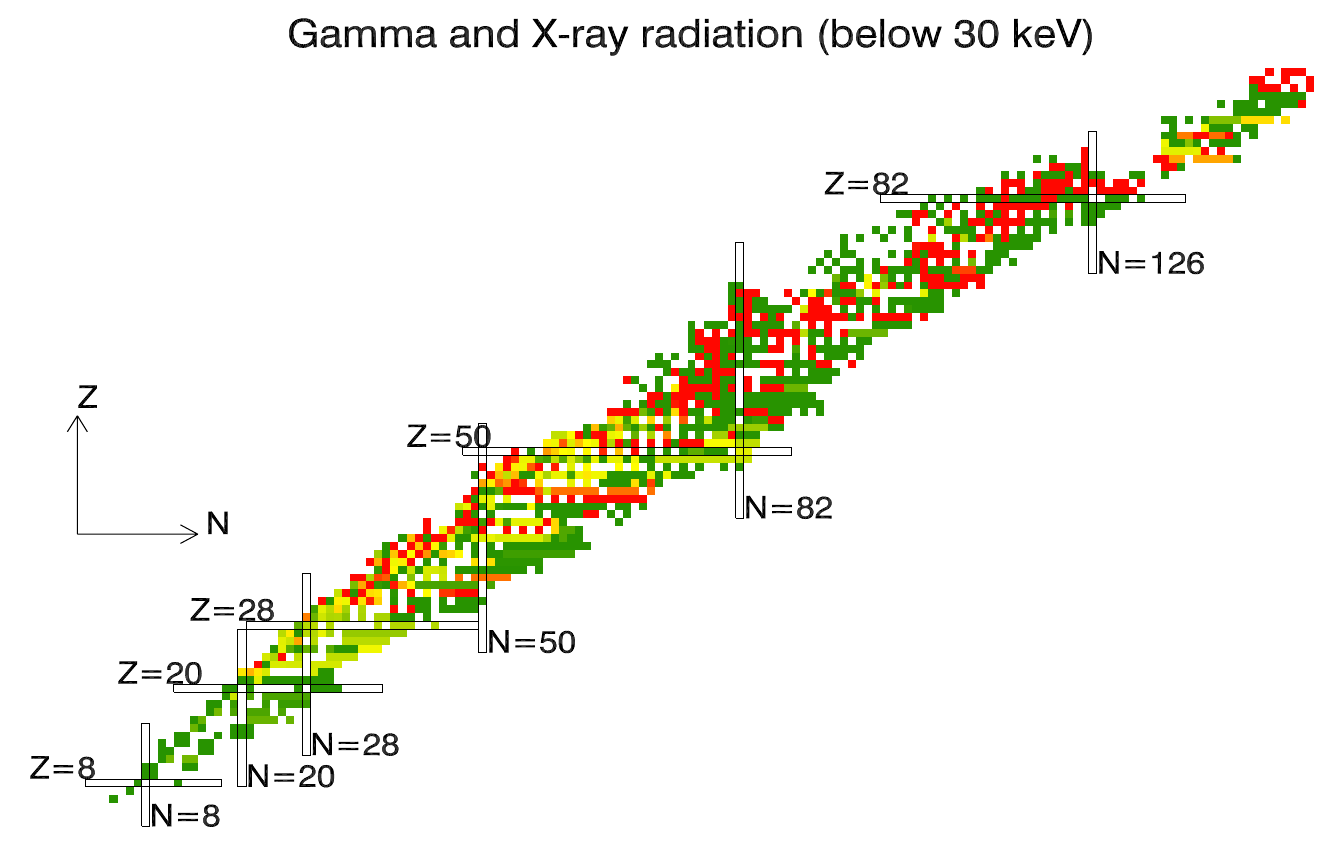}}
\hfil
\subfloat{\includegraphics[width=7.cm]{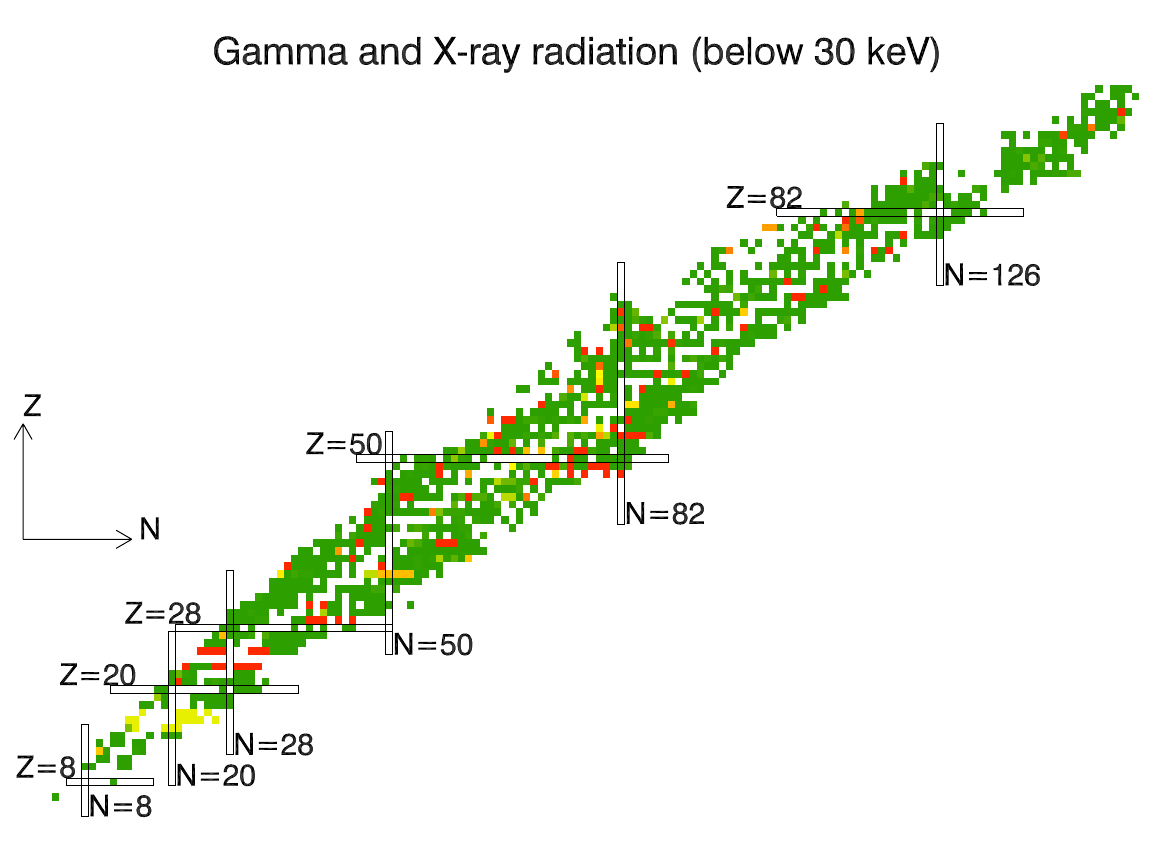}
}}
\centerline{\includegraphics[width=9cm]{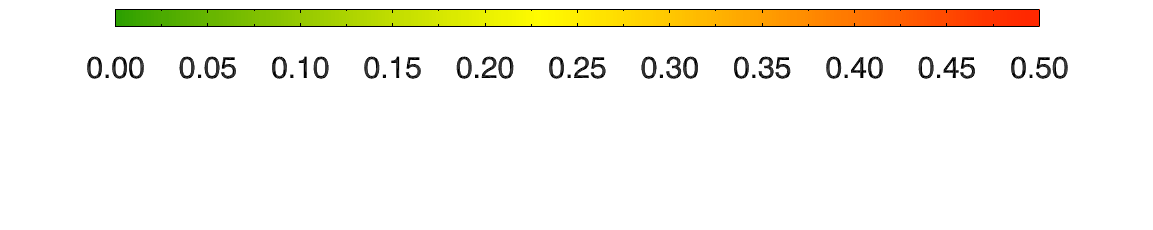}}
\vspace{-0.3in}
\caption{Nuclide charts showing the median relative intensity deviations per
isotope for X-ray emission. Simulation results using the current Geant4 code are
shown on the left; simulations using the new code are shown on the right.}
\label{fig_nuclide}
\end{figure}

Figure~\ref{fig_nuclide} shows a comparison of the performance of
the current Geant4 radioactive decay simulation with results produced by the new
code employing an improved selection of atomic parameters.
For fluorescence and Auger emission the deviation between evaluated ENSDF
intensities and those produced by current Geant4 code is large, as is apparent
from the nuclide chart shown in figure~\ref{fig_nuclide}; it amounts to $52.6\pm
2.0\%$ for X-ray emissions.
The new code yields better results, with deviations amounting to $4.1\pm1.8\%$
for X-ray emission.
This is a more than 10-fold improvement with respect to the current Geant4 code.

Further studies are in progress to optimize the selection of atomic parameters
relevant to the simulation of radioactive decay with Geant4.

\section{Combination of physics models for particle transport}

Some physics models of particle interactions with matter are applicable only to
a limited energy range; Monte Carlo codes for particle transport usually exploit
a series of different models for a given physics process to cover a wide energy
range relevant to experimental applications.

Empirical algorithms are applied in the course of the simulation to manage the
transition across different models; they are often prone to generate
inconsistencies in experimentally relevant observables resulting from the
simulation, such as discontinuities in physical distributions.

Merging offline the data on which physics processes are based is an effective
alternative to blending different implemented models at runtime.
This method consists in producing data tabulations (e.g. of cross section data),
which may derive from complex theoretical calculations, parameterization of
experimental data or semi-empirical analytical models, and are specialized for a
given energy range;
the distributions produced by different calculation models are then merged by 
means of smoothing algorithms, and tabulated in a data library.
The physics quantities needed for Monte Carlo particle transport
are obtained by interpolation of the tabulated data.

This method allows the use of a single physics model to provide functionality
for simulation over an extended energy range: it reduces the risk of
inconsistencies in the resulting physics observables, since they are produced by a
unique model, and contributes to improved computational performance, since no
algorithms for blending models at runtime are needed.
Nevertheless, it retains all the advantages of specialized physics behaviour as
a function of the interacting particle energy.

A feasibility study for the application of this method has been performed: it
concerns the production of cross sections for electron impact ionization that
cover an extended energy range, from a few electronvolts to 100~GeV. The cross
sections derive from two theoretical modeling approaches: the Deutsch-M\"{a}rk
model \cite{dm1987} at low energy, recently implemented for use with Geant4
\cite{beb_tns}, and Seltzer’s approach as adopted by EEDL (Evaluated Electron Data
Library) \cite{eedl} at higher energy.
They are merged by means of smoothing algorithms available in the R statistical
analysis toolkit \cite{R}.
Cross section tabulations based on this technique are collected into a data library
for electron impact ionization, which is being developed for use with Geant4.

\begin{figure}
\centerline{\includegraphics[angle=0,width=12cm]{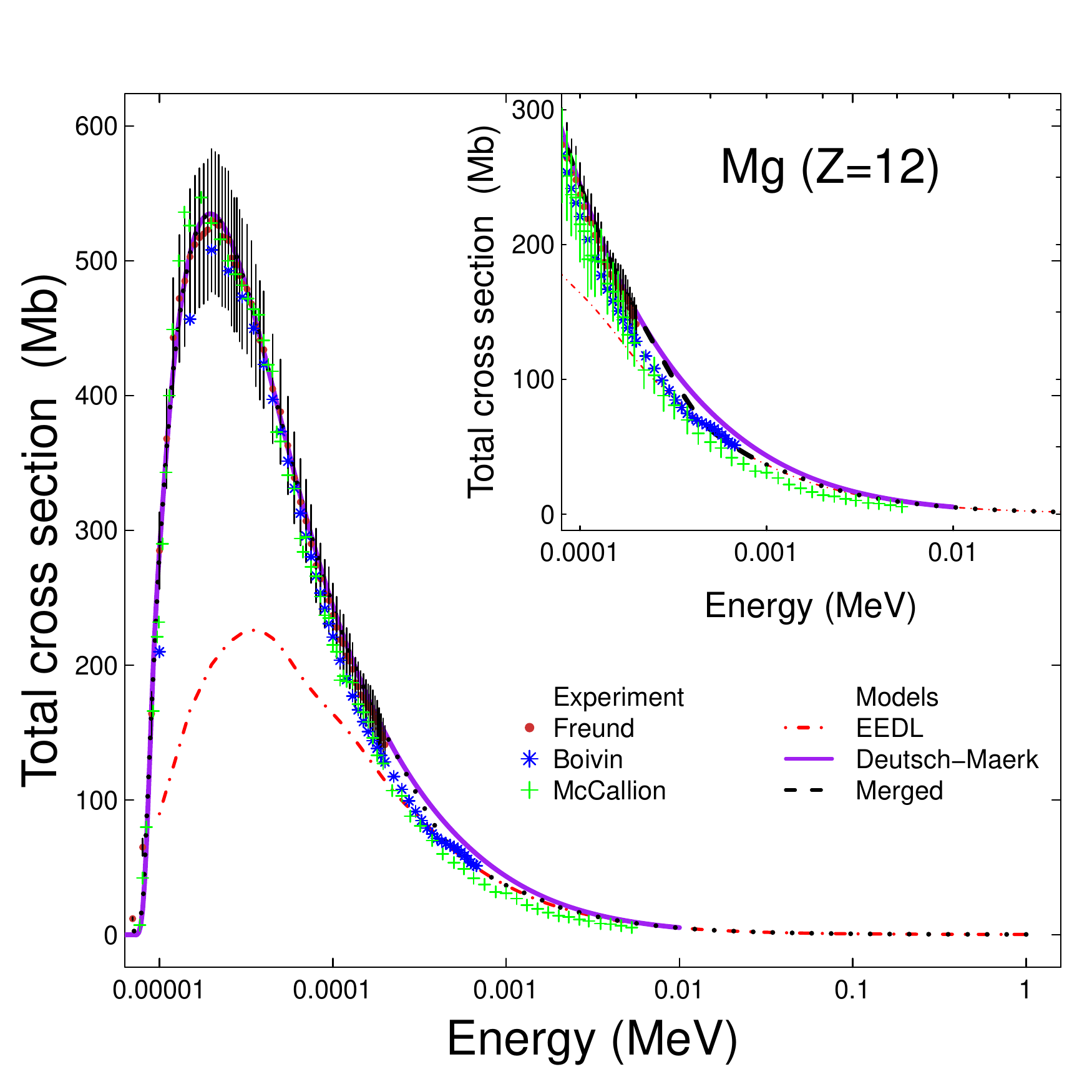}} 
\caption{An example of merging cross section data: cross sections for electron
impact ionisation of magnesium calculated by the Deutsch-M\"{a}rk model (solid
purple curve) and derived from EEDL (red dot-dashed curve) are merged (black
dashed curve) by means of a smoothing algorithm. Experimental data
\cite{freund,boivin,mccallion} support the validation of the procedure.
Experimental error bars are shown only for a subset of the experimental data for
better clarity of the figure. }
\label{fig_smooth}
\end{figure}

An example of merging ionisation cross sections calculated by two models is
shown in figure~\ref{fig_smooth}; the cross sections calculated by the two
models are merged by means of the \textit{loess} (local polynomial regression
fitting) smoothing algorithm implemented in R.

Research is in progress, with the support of experts in mathematical and
statistical techniques, to optimize the methods for merging cross section data
and other physical distributions used in Geant4 for the simulation of electron
and photon interactions.

\section*{Conclusions}

A significant effort has been invested into the evaluation and improvement of
data libraries used by Monte Carlo codes.

An extensive study of atomic binding energies has highlighted the relative merits
of different compilations and their contribution to the accuracy of simulation models
using them. 
No compilation stands out as the optimal one for any use case; rather, different 
atomic data may be needed by different physics models to optimize the accuracy of simulation.
A software package addressing this requirement has been developed.

Improvements to Geant4 radioactive decay simulation involve 
not only new algorithms, but also new sets of nuclear and atomic data.

A method to combine physics models applicable to different energy ranges by
merging data libraries, rather than blending them at runtime, has been illustrated.

\section*{Acknowledgements}
We thank the CERN Library, in particular Tullio Basaglia, for support to the research described in this paper. 
This work has been partly funded by CNPq BEX6460/10-0 grant, Brazil.



\section*{References}
\begin{thebibliography}{9}

\bibitem{g4nim}
Agostinelli S et al. 2003
\textit{Nucl. Instrum. Meth. A} {\bf 506} 250

\bibitem{g4tns}
Allison J et al. 2006
\textit{IEEE Trans. Nucl. Sci.} {\bf 53} 270

\bibitem{binding_tns}
Pia M G et al 2011
\textit{IEEE Trans. Nucl. Sci.} {\bf 58} 3246

\bibitem{bearden}
Bearden J A and Burr A F 1967
\textit{Rev. Mod. Phys.} {\bf 39}  125

\bibitem{isics}
Liu Z and Cipolla S J 1996
\textit{Comp.\ Phys.\ Comm.} {\bf 97} 315

\bibitem{carlson}
Carlson T A 1975
\textit{Photoelectron and Auger spectroscopy} Plenum, New York

\bibitem{mcnp}
X-5 Monte Carlo Team 2005
\textit{Los Alamos National Laboratory Report LA-UR-03-1987}

\bibitem{mcnpx}
Hendricks J S 2006
\textit{Los Alamos National Laboratory Report LA-UR-06-7991}

\bibitem{penelope}
Baro J et al. 1995
\emph{Nucl. Instrum. Meth. B} {\bf 100} 31

\bibitem{eadl}
Perkins S T et al 1991 
\textit{Tables and Graphs of Atomic Subshell and Relaxation Data Derived from the LLNL Evaluated 
Atomic Data Library (EADL), Z=1-100} UCRL-50400 Vol. 30 

\bibitem{sevier1979}
Sevier K 1979
\textit{Atom. Data Nucl. Data Tables} {\bf 24} 323

\bibitem{gupix}
Maxwell J A 1989
\emph{Nucl. Instrum. Meth. B} {\bf43} 218

\bibitem{toi1978}
Lederer M and Shirley V C 1978
\{textit{Table of Isotopes 7th ed.} John Wiley \& Sons, New York

\bibitem{toi1996}
Firestone R B et al. 1996
\textit{Table of Isotopes 8th ed.} John Wiley \& Sons, New York

\bibitem{egsnrc}
Kawrakow I et al. 2010 
\textit{NRCC PIRS-701} 

\bibitem{egs5}
Hirayama H et al. 2006
\textit{ SLAC-R-730 Report} Stanford, CA

\bibitem{xbook}
Thompson A C et al. 2009
\textit{X-ray Data Booklet}  Berkeley, CA, USA

\bibitem{nist_xps}
Rumble J R et al 1992
\textit{Surf. Interface Anal.} {\bf 19} 241

\bibitem{deslattes}
Deslattes R D et al. 2003
\textit{Rev. Mod. Phys.} {\bf 75} 35


\bibitem{rdm_chep2000}
Truscott P et al 200
{\textit CHEP 2000 Proceedings} A123

\bibitem{rdm_radecs2000}
Truscott P et al. 2000
{Proc. Radiation Effects Data Workshop, 2000} 147

\bibitem{ensdf}
Tuli J K 2001
{\textit Evaluated Nuclear Structure Data File: A Manual for Preparation of Datasets}
Brookhaven National Laboratory

\bibitem{Band1976433}
Band I et al. 1976
\emph{At. Data Nucl. Data Tables} {\bf 18} 433 

\bibitem{Roesel197891}
Roesel F et al. 1978 
\emph{At. Data Nucl. Data Tables} {\bf 21} 91 

\bibitem{Hager19681}
Hager R and Seltzer E 1968
\emph{Nucl. Data Sheets A} {\bf 4} 1 

\bibitem{rdm_nss2009}
Hauf S et al 2009
\textit{IEEE Nucl. Sci. Symp. Conf. Rec. } 2060

\bibitem{rdm_mc2010}
Hauf S et al. 2010
\textit{SNA+MC Proc.} Full Paper no. 10275


\bibitem{dm1987} 
Deutsch H and M\"ark D T 1987
\textit{Int. J. Mass Spectrom. Ion Processes} {\bf 79} R1

\bibitem{beb_tns}
Seo H et al. 2011
\textit{IEEE Trans. Nucl. Sci.} {\bf 58} 3219

\bibitem{eedl}
Perkins  S T et al. 1997
\textit{Tables and Graphs of Electron-Interaction Cross Sections from 10 eV to 100 GeV Derived from the 
LLNL Evaluated Electron Data Library (EEDL)} UCRL-50400 Vol. 31

\bibitem{R}
Ihaka R and Gentleman R 1996
\textit{J. Comp. Graph. Stat.} {\bf5} 299 

\bibitem{freund} 
R. S. Freund R S et al 1990
\textit{Phys. Rev. A} {\bf 41} 3575

\bibitem{boivin}
Boivin  R F and Srivastava S K 1998
\textit{J. Phys. B: At. Mol. Opt. Phys.} {\bf 31} 2381

\bibitem{mccallion} 
McCallion P et al. 1992
\textit{J. Phys. B: At. Mol. Opt. Phys.} {\bf. 25} 1051


\end{thebibliography}
\end{document}